\documentclass{elsarticle}

\usepackage{amssymb}
\usepackage{amsthm}

\newtheorem{inv}{Invariant}

\usepackage[ruled,vlined]{algorithm2e}

\begin{document}


\title{Simple DFS on the Complement of a Graph and on Partially Complemented Digraphs}


\author{Benson Joeris}
\address{Department of Combinatorics and Optimization, 
University of Waterloo, 
Waterloo, ON N2L 3G1, Canada}

\author{Nathan Lindzey}
\address{Department of Mathematics 
Colorado State University, 
Fort Collins, CO 80521}

\author{Ross M. McConnell}
\address{Department of Computer Science,
Colorado State University, 
Fort Collins, CO 80521}

\author{Nissa Osheim}
\address{Department of Computer Science,
Colorado State University, 
Fort Collins, CO 80521}

\begin{abstract}

A \emph{complementation operation} on a vertex of a digraph changes all 
outgoing arcs into non-arcs, and outgoing non-arcs
into arcs. A \emph{partially complemented digraph} $\widetilde{G}$ is a 
digraph obtained from a sequence of vertex complement operations on $G$.  
Dahlhaus et al.~showed that,
given an adjacency-list representation of $\widetilde{G}$, 
depth-first search (DFS) on $G$ can be performed in $O(n + \widetilde{m})$ time,
where $n$ is the number of vertices and $\widetilde{m}$ is the number of edges
in $\widetilde{G}$.  To achieve this 
bound, their algorithm makes use of a somewhat complicated 
stack-like data structure to simulate the recursion stack, instead of
implementing it directly as a recursive algorithm.
We give a recursive $O(n+\widetilde{m})$ 
algorithm that uses no complicated data-structures.
\end{abstract}

\maketitle

\section{Introduction}

A \emph{complementation operation} on a vertex of a digraph changes all 
outgoing arcs into non-arcs, and outgoing non-arcs
into arcs. A \emph{partially complemented digraph} $\widetilde{G}$ is a 
digraph obtained from a sequence of vertex complement operations on $G$. 
Let $n$ denote the number of vertices and $m$ denote the number of edges 
of $\widetilde{G}$. In~\cite{DahlhausGM02} several linear-time 
graph algorithms for partially complemented digraphs were presented.  
Their algorithm for DFS on partially complemented digraphs
was notably more complicated than their algorithm for BFS despite the 
comparable simplicity of DFS and BFS in the usual context.

One application they give of this result is in computing the modular
decomposition of an undirected graph $G$.  A step in the 
algorithm of~\cite{EGMS94} requires
finding the strongly-connected components of a directed graph $G'$
that is in the partially-complemented equivalence class of $G$, but
whose size is not bounded by the size of $G$.  Construction
of $G'$ and running DFS on it gives the $\Theta(n^2)$
bottleneck in the running time of that algorithm.  An $O(n+m \log n)$ bound
is easily obtained for this algorithm using the partially-complemented
DFS algorithm of~\cite{DahlhausGM02}, and they use it to obtain
a simple linear-time algorithm for modular decomposition.

Their algorithm for DFS is not recursive and is 
complicated by the use of so-called \emph{complement stacks}, a stack-like 
data structure to simultaneously simulate the recursion stack and keep 
track of which undiscovered vertices will not be called from which 
vertices on the recursion stack.  This raised the question of 
whether there exists a more natural recursive DFS algorithm for partially 
complemented digraphs.  To this end, we give an elementary recursive 
$O(n+\widetilde{m})$ algorithm for performing depth-first search on $G$ 
given a partially complemented digraph $\widetilde{G}$.

A notable special case is when every vertex is complemented, that is, 
$\widetilde{G} = \overline{G}$ where $\overline{G}$ denotes the complement 
of $G$. Algorithms for performing DFS on $G$, given $\overline{G}$, have also 
been developed~\cite{ItoY98}\cite{KaoOT98}, the most efficient of which 
runs in $O(n+\overline{m})$ time where $n + \overline{m}$ is the number 
of vertices and number edges of $\overline{G}$ respectively. To achieve 
this bound, the algorithm in~\cite{ItoY98} makes use of the Gabow-Tarjan 
disjoint set data structure~\cite{GabowT83}.  Our algorithm also provides 
a simpler way to run DFS on $G$ given $\overline{G}$.

\section{Preliminaries}

We will assume that the vertices are numbered 1 through $n$.
For vertices $u$ and $v$, let $u < v$ denote that the vertex
number of $u$ is smaller than that of $v$.

Let $\widetilde{N}(v)$ denote the neighbors of $v$ in $\widetilde{G}$.  That is, if $v$ 
is uncomplemented, $\widetilde{N}(v)$ is a list of neighbors of $v$ in $G$,
and if $v$ is complemented, it is a list of non-neighbors of $v$ in $G$.
We are given $\widetilde{N}(v)$ for each vertex
$v$.  We assume that $\widetilde{N}(v)$ is given in a doubly-linked list,
sorted by vertex number.   (This ordering can be achieved in 
$O(n+\widetilde{m})$
time by radix sorting the set $\{(v,w)| v \in V$ and $w \in \widetilde{N}(v)\}$
using $v$ as the primary sort key and $w$ as the secondary sort key.)
Each vertex is labeled with a bit to indicate whether it is complemented,
specifying whether $\widetilde{N}(v)$ should be interpreted as neighbors
or non-neighbors of $v$ in $G$.  We will find it convenient to assume that
$\widetilde{N}(v)$ is terminated by a fictitious vertex whose vertex number,
$n+1$, is larger than those of any vertex in $G$.
Let us call this the {\em pc-list representation of $G$}.

\section{The Algorithm}

\begin{algorithm}\label{alg:dfs}
\caption{DFS($v$)} 
\KwData{A current undiscovered vertex $v$, and a global ordered doubly-linked list $U$ of undiscovered vertices}
\texttt{remove}($U$,$v$)\;
\If{$v$ is uncomplemented}{
	\For{$u \in N(v)$}{
	    \If{$u$ is undiscovered}{
	 	   DFS($u$)\;
		}
	}
}
\Else{
	$u_v \longleftarrow$ \texttt{head}($U$)\;
	$n_v \longleftarrow$ \texttt{head}($\widetilde{N}(v)$)\;
	\While{$u_v \neq null$}{
	\If{$u_v = n_v$}{
		$u_v \longleftarrow$ \texttt{next}($U,u_v$)\;
		$n_v \longleftarrow$ \texttt{next}($\widetilde{N}(v),n_v$)\;
	}
	\ElseIf{$u_v > n_v$}{
		$s \longleftarrow n_v$\;
		$n_v \longleftarrow$ \texttt{next}($\widetilde{N}(v),n_v$)\;
		\texttt{remove}($\widetilde{N}(v),s$)\;
	}
\Else{
	DFS($u_v$)\;
	\tcp{restarting step ...}
        $w$ = prev($\widetilde{N}(v),n_v$)\;
	\While{$w \neq null$ and $w \notin U$}{
		$t = w$\;
		$w \longleftarrow$ \texttt{prev}($\widetilde{N}(v),t$)\;	
		\texttt{remove}($\widetilde{N}(v), t$))\;
	}
	\If{$w = null$}{
		$u_v \longleftarrow$ \texttt{head}($U$)\;
	}
	\Else{
		$u_v \longleftarrow$ \texttt{next}($U,w$)\;
	}
}
}
}

\end{algorithm}

A vertex is {\em discovered} when a recursive call to DFS is made on it.
At all times, we maintain a doubly-linked list $U$ of undiscovered vertices, which
is sorted by vertex number.  Initially, $U$ contains all vertices of $G$.

When a recursive call is made on an undiscovered vertex $v$, it
is removed from $U$ and marked as discovered.  
If $v$ is uncomplemented,
the algorithm for generating recursive calls from it is exactly
what it is in standard DFS:  for each $w \in \widetilde{N}(v)$,
if $w$ is undiscovered, a recursive call is made on it.

If $v$ is complemented, then the presence of each $w \in \widetilde{N}(v)$
is used to block any recursive call on $w$ from $v$, since
this means that $w$ is not a neighbor of $v$
in $G$.  If $w$ has been discovered, however, 
its absence from $U$ suffices to block a recursive call on it from $v$.
This allows us to remove $w$ from $\widetilde{N}(v)$ while maintaining
the following invariant:

\begin{inv}\label{inv:remove}
All undiscovered non-neighbors of each complemented vertex $v$ remain in $\widetilde{N}(v)$.
\end{inv}

The invariant suffices to prevent recursive calls from $v$ on non-neighbors
of $v$.  Removal of elements from $\widetilde{N}(v)$ while maintaining
this invariant is the key to our time bound, since it may be necessary 
to traverse an element $w \in \widetilde{N}(v)$ more than once, and
we can charge the extra cost of multiple traversals to deletions of vertices
from $\widetilde{N}(v)$.

We traverse the lists for $\widetilde{N}(v)$ and $U$ in parallel,
in a manner similar to the {\tt merge} operation in {\tt mergesort},
advancing the pointer to the lower-numbered vertex at each step,
or advancing both pointers in their lists if they point to the same
vertex.

Let $n_v$ be the current vertex in $\widetilde{N}(v)$ and let $u_v$
be the current vertex in $U$.
If $n_v = u_v$, it is a non-neighbor of $v$, hence we cannot
make a recursive call on it from $v$.   
We set $n_v$ to its successor in $\widetilde{N}(v)$ and $u_v$ to its successor
in $U$.  If $n_v \not\in U$, which is detected if $n_v < u_v$,
then $n_v$ has already been discovered, and we can remove remove it from
$\widetilde{N}(v)$ and advance $n_v$ to the next vertex in
$\widetilde{N}(v)$ by Invariant~\ref{inv:remove}.
If $u_v \not\in \widetilde{N}(v)$, which
is detected if $u_v < n_v$, we make a recursive call on $u_v$,
and when this recursive call
returns, we perform the following {\em restarting step}:

\begin{itemize}
\item Advance $u_v$ to be the first vertex $u'_v$ in $U$ with a higher vertex number
than the current $u_v$.
\end{itemize}

The difficulty in implementing the restarting step efficiently is that
when the recursive call on $u_v$ returns, $u_v$ and possibly many other
vertices were discovered and removed from $U$ during the recursive call on it.
It is thus not a simple
matter of finding the successor of $u_v$ in $U$.  We discuss an efficient
implementation below.

By induction on the number of times $u_v$ is advanced, we see
that the following invariant is maintained:

\begin{inv}\label{inv:predecessors}
Whenever $u_v$ advances in $U$, its predecessors in $U$ are members
of $\widetilde{N}(v)$, hence non-neighbors of $v$.
\end{inv}

The call on $v$ returns when $u_v$ moves past the end of $U$, which
happens before $n_v$ moves past the end of $\widetilde{N}(v)$, due
to the presence of the fictitious vertex numbered $n+1$ at the end
of $\widetilde{N}(v)$.

For the correctness, let $k$ be the number of undiscovered vertices
when $v$ is discovered.  Since the number of undiscovered vertices
is less than $k$ when each recursive call is generated from $v$,
we may assume by induction on the number of undiscovered vertices
that each recursive call generated from $v$ faithfully executes a DFS,
given the marking of vertices as undiscovered or discovered when the call is made.
If $v$ is not complemented, the correctness of the call on it is
immediate.  If $v$ is complemented, the correctness of the call on it
follows from Invariant~\ref{inv:predecessors}
and the fact that a recursive call is made on $u_v$ whenever it is found
to be a neighbor of $v$.

To implement the restarting step,
we traverse $\widetilde{N}(v)$ backward, starting
at $\texttt{prev}(\widetilde{N}(v),n_v)$, removing vertices 
from $\widetilde{N}(v)$ that are no longer
in $U$.  Their removal does not violate Invariant~\ref{inv:remove} or
Invariant~\ref{inv:predecessors}.

Eventually, we have either encountered a vertex $w$ that is in both $\widetilde{N}(v)$ and $U$,
or we have deleted all precessors of $u_v$ in $\widetilde{N}(v)$.  Suppose we encounter $w$.
All predecessors of $u_v$ in $U$ were non-neighbors of $v$ when $n_v$
was last advanced.  Also, $u_v < n_v$, since we wouldn't have reached
$u_v$ to make a recursive call on it if $n_v$ were less than $u_v$, and
we wouldn't have made a recursive call on it if they were equal.
Finally, $w$ is now the predecessor
of $n_v$ in $\widetilde{N}(v)$ since we have deleted all vertices
from $n_v$ back to $w$.  It follows that the successor of $w$ is greater
than $u_v$, hence the successor of $w$ now gives $u'_v$.
By a similar argument, if all predecessors of $n_v$ are deleted from $\widetilde{N}(v)$,
the first element of $U$ gives $u'_v$.

Though the restarting step is not an $O(1)$ operation, the total time required by restarting
steps over the entire DFS is $O(n + \widetilde{m})$, since all but $O(1)$ of
a restarting operation can be charged to elements that it deletes from some list $\widetilde{N}(v)$,
and the initial sum of sizes of these lists is $\widetilde{m}$.
Pseudocode of the algorithm is given as Algorithm~\ref{alg:dfs}, and makes use
of the following $O(1)$ operations:

\begin{itemize}
\item \texttt{head}($L$): returns the head node of a doubly-linked list $L$.
\item \texttt{next}($L$,$n$): returns the next node of the node $n$ that exists in $L$, or null if no such node exists.
\item \texttt{prev}($L$,$n$): returns the previous node of the node $n$ that exists in $L$, or null if no such node exists.
\item \texttt{remove}($L$,$n$):  removes a node $n$ from doubly-linked list $L$.
\end{itemize}

\bibliographystyle{plain}
\bibliography{master}
\end{document}